\documentclass[12pt]{article}
\usepackage{bbm}
\usepackage{amssymb}
\usepackage{pifont}
\usepackage{mathrsfs}
\usepackage{amsfonts}
\usepackage{color}
\usepackage{footmisc}
\usepackage[numbers,sort&compress]{natbib}
\usepackage[colorlinks,linkcolor=blue,citecolor=blue,urlcolor=blue]{hyperref}

\newcommand{\omits}[1]{}

\begin{document}

\begin{center}
{\bf \LARGE de Sitter-covariant Hamiltonian formalism\bigskip

of Einstein--Cartan gravity}
\bigskip\bigskip

{\large Jia-An Lu\footnote{Email: ljagdgz@163.com}}
\bigskip

School of Physics, Sun Yat-sen University,\\ Guangzhou 510275, China
\bigskip

\begin{abstract}
The Hamiltonian formalism of Einstein--Cartan (EC) gravity is a starting point
for canonical quantum gravity. The existing formalisms are at most Lorentz
covariant, or diffeomorphism covariant. Here we analyze the Hamiltonian EC
gravity in a 5d covariant way, with the gauge group being the de Sitter (dS)
group, which unifies the Lorentz transformations and translation in an
elegant manner, and also coincides with the acceleration of the universe.
We reformulate the EC equations into a dS-covariant form, then find out the
dS-covariant constraints of the phase space, and make all the constraint
functions constitute a closed algebra by constructing a dS-invariant Dirac
bracket, for the purpose of quantization.
\end{abstract}
\end{center}



\section{Introduction}
In search of a quantum theory of gravity, generally one should start from a classical
theory and then quantize it. As the simplest classical theory of gravity, Einstein's
general relativity (GR) is unnatural in the viewpoint of a gauge theory, since the
Lorentz connection as the gauge potential is not an independent variable. By including
spacetime torsion, the Lorentz connection becomes independent, then one obtains the
Einstein--Cartan (EC) theory of gravity \cite{EC}.

The EC gravity is usually interpreted as a Poincar\'e gauge theory of gravity, in
which the gauge transformations are the Lorentz and diffeomorphism transformations,
acting on the Lorentz connection and co-tetrad field \cite{Kibble,Hehl76}. But there
exist alternative interpretations, where the Lorentz connection and co-tetrad field
are combined into a 5d connection, valued at the Poincar\'e/de Sitter/anti-de
Sitter (P/dS/AdS) algebra \cite{Guo76,West80,Pilch,Tseytlin}. Then the gauge
transformations consist of the P/dS/AdS and diffeomorphism transformations, acting
on the 5d connection and a 5d vector field $\xi^A$. Actually, $\xi^A$
constitutes a system of local 5d Minkowski coordinates, named the local inertial
coordinates (LIC) \cite{West80,Lu17}.

In this formulation, there exist some gauges in which the 5d connection reduces to
the Lorentz connection and co-tetrad field. These gauges constitute a Lorentz subgroup
of the P/dS/AdS symmetry. Also, matter fields in the standard model of particle physics
are described by representations of the Lorentz group, other than the complete
P/dS/AdS group. For these reasons, it is argued that the 5d connection should be projected
into a Cartan connection, with the Lorentz group as the stability subgroup
\cite{West80,Tseytlin,Wise}. Note that the Cartan connection transforms nonlinearly
under the complete gauge group, and so the corresponding formalism is called nonlinear
realization \cite{Tresguerres}.

However, the P/dS/AdS symmetry remains the true symmetry for gravitation. Accordingly,
the original linearly realized formulation \cite{Kawai,Grignani,Lu13,Lu14}
is more fundamental
than the nonlinear realization. Besides, the linear realization is able to include
those matter fields transforming under the full representations of the P/dS/AdS
group \cite{Kazmierczak,Takook}. Such new matter fields might be responsible for
new physics \cite{Aldrovandi,Lu16}.

Moreover, among the P, dS and AdS groups, the dS group seems to be the best
choice for gravity. Firstly, the dS/AdS group provides an elegant unification of
the Lorentz transformations and translation, resulting in a 5d covariant theory.
For example, the energy-momentum conservation and angular momentum conservation
can be united into a 5d covariant conservation in the dS/AdS gravity \cite{Lu16}.
Secondly, the dS group is consistent with the asymptotic symmetry of the
expanding universe.

In this paper, the EC gravity is analyzed in the linearly realized formalism, with
the dS group as the gauge group. It is shown that the field equations, consisting
of the Einstein equation and Cartan equation, can be united into a 5d covariant
equation. Then we go on to the Hamiltonian formalism. Making use of the Legendre
transformation, the configuration tangent bundle is transformed into the phase
space. Next, with the help of the Arnowitt--Deser--Misner (ADM) decomposition,
we find out the first class constraints which generate the dS and diffeomorphism
transformations on the constraint surface. Furthermore, the Poisson bracket is
modified into a Dirac bracket, such that all the second class constraints become
first class, and so all the constraint functions form a closed Poisson algebra.
The work paves the way for the canonical quantization of the dS-covariant
theory of gravity.

The paper is organized as follows. In section 2, the EC gravity is formulated in a
dS-covariant way. In section 3, the Hamiltonian formalism of the theory is calculated.
In section 4, we give some remarks on the linear realization.

Here are the conventions to be used. The Greek letters $\mu,\nu\ldots$ label the
spacetime indices and take the values $t,a$, being lowered or raised by the
metric $g_{\mu\nu}$ or its inverse. The Latin letters $a,b\ldots$ label the spatial
indices and run over $x,y,z$. Also, the Latin letters $A,B\ldots$ refer to the $SO(1,4)$
indices and take the values $0,1\cdots4$, being lowered or raised by
$\eta_{AB}={\rm diag}(-1,1\cdots1)$ or its inverse. The Greek letters $\alpha,\beta\ldots$
refer to the $SO(1,3)$ indices and run over $0,1\cdots3$, being lowered or raised by
$\eta_{\alpha\beta}={\rm diag}(-1,1\cdots1)$ or its inverse.

\section{EC gravity as dS gravity}
\subsection{dS gravity from gauge principle}
The dS gravity is a gauge theory of the dS group. In the gauge theory, a global
symmetry is localized by introducing a gauge field. For the present case, the
global symmetry is the dS group $SO(1,4)$. Let us start from a classical matter
field with both the global dS invariance and the diffeomorphism invariance.
Its action integral reads
\begin{equation}\label{SM}
S_M=\int_\Omega d^4x\,\mathscr{L}_M\sqrt{-g},\ \
\mathscr{L}_M=\mathscr{L}_M(\psi, \partial_\mu\psi, c.c., \xi^A,
\partial_\mu \xi^A),
\end{equation}
where $\Omega$ is an arbitrary domain of the dS spacetime ${\cal M}_l$,
$\{x^\mu\}$ is an arbitrary coordinate system on $\Omega$, $g$ is the
determinant of the dS metric $g_{\mu\nu}$, $\psi$ is the matter field,
and $\xi^A$ is the radius vector field of ${\cal M}_l$, viewed in the
5d ambient Minkowski space, and subject to the condition $\eta_{AB}\xi^A
\xi^B=l^2$. Note that $g_{\mu\nu}$ is considered as a functional of $\xi^A$:
\begin{equation}\label{gdS}
g_{\mu\nu}=\eta_{AB}(\partial_\mu\xi^A)(\partial_\nu\xi^B),
\end{equation}
and so $S_M$ is a functional of $\psi$ and $\xi^A$. The conservation law
with respect to the dS and diffeomorphism symmetries of this theory is
discussed in Ref. \cite{Lu-SR}. In order to localize the dS symmetry,
introduce a dS connection $\Omega^A{}_{B\mu}$ and change the ordinary
derivative $\partial_\mu$ to be a covariant derivative $D_\mu$, e.g.,
$D_\mu\xi^A=\partial_\mu\xi^A+\Omega^A{}_{B\mu}\xi^B$. It follows that
\cite{Guo76,West80}
\begin{equation}\label{metric}
g_{\mu\nu}=\eta_{AB}(D_\mu\xi^A)(D_\nu\xi^B).
\end{equation}
Consider the gauges with $\xi^\alpha=0,\xi^4=l$. For any dS transformation
given by the group element $h^A{}_B\in SO(1,4)$, $\xi^A$ transforms to
$h^A{}_B\xi^B$. To preserve the gauge condition, there should be $h^\alpha{}
_{\beta}\in SO(1,3)$, $h^\alpha{}_4=h^4{}_\alpha=0$, and $h^4{}_4=1$. For this
reason, we call the gauges the Lorentz gauges. Then Eq. (\ref{metric})
reduces to $g_{\mu\nu}=\eta_{\alpha\beta}(D_\mu\xi^\alpha)(D_\nu\xi^\beta)$,
implying that $D_\mu\xi^\alpha$ is an orthonormal co-tetrad field, denoted
by $e^\alpha{}_\mu$. Moreover, note that $D_\mu\xi^\alpha=\Omega^\alpha{}_{4\mu}
\cdot l$, and thus $\Omega^\alpha{}_{4\mu}=l^{-1}e^\alpha{}_\mu$. Also, the
geometrical meaning of $\Omega^\alpha{}_{\beta\mu}$ can be read off from its
transformation property: it is just a Lorentz connection, denoted by $\Gamma
^\alpha{}_{\beta\mu}$. In conclusion, in the Lorentz gauges, the dS connection
\cite{Guo76,Tseytlin}
\begin{equation}\label{Omega}
\Omega^{AB}{}_{\mu}=\left(
\begin{array}{cc}
\Gamma^{\alpha\beta}{}_{\mu}&l^{-1}e^{\alpha}{}_\mu\\
-l^{-1}e^\beta{}_\mu&0
\end{array}
\right).
\end{equation}
It is derived from the gauge principle, other than being defined ad hoc.
To complete the construction of dS gravity, introduce the action integral of the
gravitational field:
\begin{equation}\label{SG}
S_G=\int_\Omega d^4x\,\mathscr{L}_G\sqrt{-g},\ \
\mathscr{L}_G=\mathscr{L}_G(\xi^A, D_\mu\xi^A, {\cal F}^{AB}{}_{\mu\nu}),
\end{equation}
where ${\cal F}^A{}_{B\mu\nu}=d_\mu\Omega^A{}_{B\nu}+\Omega^A{}_{C\mu}
\wedge\Omega^C{}_{B\nu}$ is the dS curvature. Define $S=S_M+\kappa S_G$
and the variational derivatives
$V_{AB}{}^\mu,V_A$ by $\delta S=\int_\Omega
d^4x\,(V_{AB}{}^\mu\,\delta\Omega^{AB}{}_\mu+V_A\,\delta\xi^A)\sqrt{-g}$,
where $\kappa$ is the coupling constant, and $V_A\,\xi^A\equiv0$, because
$\delta\xi^A$ is constrained by $\xi^A\xi_A=l^2$. Then the gravitational
field equations consist of $V_{AB}{}^\mu=0$ and $V_A=0$. Also, the conservation
law with respect to the local dS symmetry and diffeomorphism symmetry is
discussed in Ref. \cite{Lu16}. With the help of this, we have
\begin{equation}\label{VAB}
V_{AB}{}^\mu=\tau_{AB}{}^\mu+\Sigma_\nu{}^\mu D^\nu\xi_{[A}\cdot\xi_{B]},
\end{equation}
\begin{equation}\label{VA}
V_A=V_{BC}{}^{\nu}D^\mu\xi_A\cdot{\cal F}^{BC}{}_{\nu\mu},
\end{equation}
where $\tau_{AB}{}^\mu=(\partial\mathscr{L}/\partial D_\mu\psi)T_{AB}\psi+c.c.
+2D_\nu\partial\mathscr{L}/\partial{\cal F}^{AB}{}_{\mu\nu}$ is the dS spin
current, $\Sigma_\nu{}^\mu=-(\partial\mathscr{L}/\partial D_\mu\psi)D_\nu\psi+c.c.
-2(\partial\mathscr{L}/\partial{\cal F}^{AB}{}_{\mu\sigma}){\cal F}^{AB}{}_{\nu\sigma}
+\mathscr{L}\delta^\mu{}_\nu$ is the energy-momentum tensor, $\mathscr{L}=\mathscr{L}_M
+\kappa\mathscr{L}_G$, and $T_{AB}$ are representations of the dS
generators. In the special relativity limit with ${\cal F}^{AB}{}_{\mu\nu}=0$,
$V_A=0$ holds automatically. In
the general theory of dS gravity, $V_A=0$ as long as $V_{AB}{}^\mu=0$.
Hence, the gravitational field equation is only given by $V_{AB}{}^\mu\equiv\delta S
/\delta\Omega^{AB}{}_\mu=0$.

\subsection{EC theory of gravity revisited}
So far the gravitational Lagrangian function (\ref{SG}) is rather arbitrary.
To recover the EC gravity (with a cosmological constant), put $\mathscr{L}_G
=R-2\Lambda$, where $R=R^{\alpha\beta}{}_{\mu\nu}
e_\alpha{}^\mu e_\beta{}^\nu$ is the trace of the Lorentz curvature $R^\alpha
{}_{\beta\mu\nu}=d_\mu\Gamma^\alpha{}_{\beta\nu}+\Gamma^\alpha{}_{\gamma\mu}
\wedge\Gamma^\gamma{}_{\beta\nu}$, and $\Lambda=3/l^2$. Making use of Eq.
(\ref{Omega}), in the Lorentz gauges,
\begin{equation}\label{F}
\mathcal {F}^{AB}{}_{\mu\nu}=\left(
\begin{array}{cc}
R^{\alpha\beta}{}_{\mu\nu}-l^{-2}e^{\alpha}{}_\mu\wedge e^\beta{}_\nu
&l^{-1}S^{\alpha}{}_{\mu\nu}\\
-l^{-1}S^\beta{}_{\mu\nu}&0
\end{array}
\right)
\end{equation}
(c.f. \cite{Guo76,Tseytlin}), where $S^\alpha{}_{\mu\nu}=d_\mu e^\alpha{}_\nu
+\Gamma^\alpha{}_{\beta\mu}\wedge e^\beta{}_\nu$ is the torsion 1-form. Then
the EC Lagrangian function can be rewritten by
\begin{equation}\label{EC}
R-2\Lambda={\cal F}^{AB}{}_{\mu\nu}(D^\mu\xi_A)(D^\nu\xi_B)+2\Lambda,
\end{equation}
which is dS invariant, and so valid in any gauge. Now the field equation reads
\begin{equation}\label{ECeq}
G^\mu{}_\nu D^\nu\xi_A\wedge\xi_B-D_\nu(D^\mu\xi_A\wedge D^\nu\xi_B)=\kappa^{-1}J_{AB}{}^\mu,
\end{equation}
where the first term on the left hand side corresponds to the 5d (orbital) angular
momentum current, containing the Einstein tensor $G_{\mu\nu}=R_{\mu\nu}-\frac 1 2 Rg_{\mu\nu}$;
the second term corresponds to the 5d spin current, containing the torsion and the cosmological
constant; and $J_{AB}{}^\mu=\delta S_M/\delta\Omega^{AB}{}_\mu$ is the material current.
Moreover, define the
effective energy-momentum tensor ${\cal E}_\nu{}^\mu=J_{AB}{}^\mu(D_\nu\xi^A)
(2\xi^B/l^2)$, and the spin tensor $\tau_{\nu\sigma}{}^\mu=J_{AB}{}^\mu
(D_\nu\xi^A)(D_\sigma\xi^B)$. Then the standard form of the EC equations \cite{EC,Kibble,Hehl76}
can be recovered from Eq. (\ref{ECeq}): $R^\mu{}_\nu-\frac 1 2 R\delta^\mu{}_\nu
+\Lambda\delta^\mu{}_\nu=(2\kappa)^{-1}{\cal E}_\nu{}^\mu$, $S^\mu{}_{\nu\sigma}
+2\,\delta^\mu{}_{[\nu}S^\rho{}_{\sigma]\rho}=-\kappa^{-1}\tau_{\nu\sigma}{}^\mu$.

\section{dS-covariant Hamiltonian formalism}
\subsection{Consistent constraint surface}
To perform the Hamiltonian analysis, suppose that the spacetime region $\Omega$
has a 3+1 decomposition $\Omega=\Sigma\times\mathbb{I}$, where $\Sigma$ is a
spacelike submanifold, and $\mathbb{I}$ is an open interval on the real line.
Define the Lagrangian functional $L=\int_\Sigma d^3x\,{\cal L}[q,v]$, where the
Lagrangian density ${\cal L}=(R-2\Lambda)\sqrt{-g}$, the configuration
$q=y^\mu,\Omega^{AB}{}_\mu$, and the velocity $v=\dot{q}\equiv\partial q/\partial t$.
Here $y^\mu=y^\mu(x)$ is a parametrization of the constrained vector field $\xi^A(x)$,
such that $\xi^A(x)=\xi^A(y^\mu(x))$.

Moreover, one can calculate $\delta L/\delta
\dot{y}^\mu=-2G^{t\nu}D_\nu\xi_A(\partial\xi^A/\partial y^\mu)\sqrt{-g}$,
and $\delta L/\delta\dot{\Omega}^{AB}{}_\mu=(D^t\xi_A\wedge D^\mu\xi_B)\sqrt{-g}$.
Although not explicit, it can be shown that neither of them
depends on $v$, leading to two primary constraints:
\begin{equation}\label{pcA}
\phi_\mu=\pi_\mu+2G^{t\nu}D_\nu\xi_A(\partial\xi^A/\partial y^\mu)\sqrt{-g},
\end{equation}
\begin{equation}\label{pcB}
\phi_{AB}{}^\mu=\pi_{AB}{}^\mu-(D^t\xi_A\wedge D^\mu\xi_B)\sqrt{-g},
\end{equation}
where the momenta $p=\pi_\mu,\pi_{AB}{}^\mu$ are viewed as new variables.
Then the Hamiltonian functional can be written down: $H=\int_\Sigma d^3x\,{\cal H}
[q,p,v]$, in which the Hamiltonian density ${\cal H}=p\cdot v-{\cal L}$. The consistency
condition of a constraint is that its evolution according to the Hamiltonian equations
$\dot{q}=\delta H/\delta p$ and $\dot{p}=-\delta H/\delta q$ is equal to zero.
For the primary constraints (\ref{pcA})--(\ref{pcB}), the consistency conditions
lead to two things: The first is the solution of $\dot{\Omega}^{AB}{}_a$, as a functional
of $q$ and $\dot{y}^\mu$; and the second is the secondary constraint $C_{AB}$. The results
are as below:
\begin{eqnarray}\label{Odot}
\dot{\Omega}^{AB}{}_a=D_a\Omega^{AB}{}_t-l^{-2}D_t\xi^A\wedge D_a\xi^B
+R_{tabc}(D^b\xi^A)(D^c\xi^B)\nonumber\\
+(g^{tt})^{-1}D^t\xi^A\wedge D^b\xi^B(\Lambda\,g_{ab}-R^c{}_{acb}
-R_{tbca}\,g^{ct}),
\end{eqnarray}
\begin{eqnarray}\label{sc}
C_{AB}=-(G^t{}_\nu+\Lambda\,\delta^t{}_\nu)D^\nu\xi_A\wedge\xi_B
\sqrt{-g}+T^t{}_{\nu\sigma}(D^\nu\xi_A)(D^\sigma\xi_B)\sqrt{-g},
\end{eqnarray}
where $T^\sigma{}_{\mu\nu}=S^\sigma{}_{\mu\nu}+2\delta^\sigma{}_{[\mu}S_{\nu]}$,
$S_\sigma=S^\nu{}_{\sigma\nu}$. In virtue of the Bianchi identity $D_{[\sigma}
{\cal F}^{AB}{}_{\mu\nu]}=0$, it can be verified that the secondary constraint is
consistent already. Consequently, the consistent constraint surface of the phase
space ${\cal P}$ is given by the vanishing of the constraints
(\ref{pcA})--(\ref{pcB}) and (\ref{sc}).

\subsection{First-class constraints and symmetries}
Next, we shall recombine the above constraints into two classes. A function $F[q,p]$ of the
constrained
phase space is called first class, if for any constraint $\phi$, the Poisson bracket $\{F,
\phi\}\approx 0$, i.e., $\{F,\phi\}$ vanishes on the constraint surface. Otherwise, $F$ is
called second class. Here the Poisson bracket is defined by these fundamental relations:
\begin{equation}
\{y^\mu(\vec{x}),\pi_\nu(\vec{z})\}=\delta^\mu{}_\nu\,\delta(\vec{x}-\vec{z}),
\end{equation}
\begin{equation}
\{\Omega^{AB}{}_\mu(\vec{x}),\pi_{CD}{}^\nu(\vec{z})\}=\delta^{[A}{}_C\,\delta^{B]}{}_D\,
\delta^\nu{}_\mu\,\delta(\vec{x}-\vec{z}),
\end{equation}
where $\vec{x},\vec{z}$ denote the points on the spatial surface $\Sigma$.

The first class constraints can be obtained by analyzing the first-class Hamiltonian,
which is defined by inserting the velocity solution (\ref{Odot}) into the original
Hamiltonian: $H_1[q,p,v_1]\equiv H[q,p,v_1,v_2[q,p,v_1]]$, where $v_1=\dot{y}
^\mu, \dot{\Omega}^{AB}{}_t$ are the unsolvable velocities as free parameters, and
$v_2=\dot{\Omega}^{AB}{}_a$ is the solvable velocity with the solution $v_2[q,p,v_1]
=v_2[q,v_1]$ given by Eq. (\ref{Odot}). The consistency of any constraint $\phi$
implies that $\{\phi, H_1\}\approx 0$, and so $H_1$ is a first-class function
of ${\cal P}$. Actually, $H_1$ is a first-class constraint. To see this,
first notice that
\begin{equation}\label{H1}
H_1[q,p,v_1]=H[q,p]+\int_\Sigma d^3x\,(\phi_1\,v_1+\phi_2\,v_2[q,v_1]),
\end{equation}
where $H[q,p]\equiv H[q,p,0]$, $\phi_1=\phi_\mu,\phi_{AB}{}^t$, and $\phi_2=\phi_{AB}{}^a$.
Then it suffices to show that $H[q,p]\approx 0$. By definition, $H[q,p]=-L[q,0]$.
With the help of the Noether identity $(\partial{\cal L}/\partial D_\mu\xi^A)
D_\nu\xi^A+2(\partial{\cal L}/\partial{\cal F}^{AB}{}_{\mu\sigma}){\cal F}^
{AB}{}_{\nu\sigma}={\cal L}\,\delta^\mu{}_\nu$ with respect to the diffeomorphism
invariance \cite{Lu16}, it follows that $L[q,0]=\int_\Sigma\,d^3x~\delta{L}/\delta
\Omega^{AB}{}_t\cdot\Omega^{AB}{}_t$. Moreover, in virtue of $C_{AB}=\delta L/\delta
\Omega^{AB}{}_t$, we have
\begin{equation}\label{Hpq}
H[q,p]=-\int_\Sigma d^3x\,C_{AB}\,\Omega^{AB}{}_t\approx0,
\end{equation}
and hence $H_1$ is a first-class constraint.
In $H_1$ given by Eqs. (\ref{H1})--(\ref{Hpq}), there are two sets of free parameters:
$\Omega^{AB}{}_t$ and $v_1$, which correspond to two sets of first-class constraints. To
find out these constraints, it is convenient to use the ADM decomposition \cite{ADM} of the
time direction basis vector: $(\partial_t)^\mu=Nn^\mu+N^\mu$, where $N$ is named the lapse
function, $N^\mu$ is named the shift vector, which is tangent to $\Sigma$, and $n^\mu$ is
normal to $\Sigma$, with $n^\mu n_\mu=-1$. Accordingly, $D_t\xi^A$ can be decomposed as
\begin{equation}
D_t\xi^A=ND_\perp\xi^A+N^aD_a\xi^A,
\end{equation}
where $D_\perp\xi^A=n^\mu D_\mu\xi^A$ can be solved as the functions of $D_a\xi^A$.
Note that $\dot{y}^\mu=\partial y^\mu/\partial\xi^A\cdot\dot{\xi}^A=(\partial y^\mu
/\partial\xi^A)(ND_\perp\xi^A+N^aD_a\xi^A-\Omega^A{}_{Bt}\xi^B)$, and so the free
parameters $\dot{y}^\mu$ can be replaced by $N$ and $N^a$. Putting this replacement
into Eq. (\ref{H1}) results in $H_1[q,p,v_1]=\int_\Sigma\,d^3x\,(N\tilde{{\cal H}}
_\perp+N^d\tilde{{\cal H}}_d+\dot{\Omega}^{AB}{}_t\,\phi_{AB}{}^t+\Omega^{AB}{}_t
\tilde{{\cal H}}_{AB})$,
where $\tilde{{\cal H}}_\perp=\phi_A D_\perp\xi^A+\phi_{AB}{}^a{\cal F}^{AB}{}_
{\perp a}$, $\tilde{{\cal H}}_d=\phi_A D_d\xi^A+\phi_{AB}{}^a{\cal F}^{AB}{}_{da}$,
$\tilde{{\cal H}}_{AB}=-D_a\pi_{AB}{}^a-\pi_{[A}\,\xi_{B]}$, ${\cal F}^{AB}{}_
{\perp a}={\cal F}^{AB}{}_{\mu a}n^\mu$, $\pi_A=\pi_\mu\,\partial y^\mu/\partial
\xi^A$, and likewise, $\phi_A=\phi_\mu\,\partial y^\mu/\partial\xi^A$.

It can be shown that the constraints $\tilde{{\cal H}}_\perp, \tilde{{\cal H}}_d,
\tilde{{\cal H}}_{AB}$, and $\phi_{AB}{}^t$ do not depend on the free parameters
$N, N^d, \dot{\Omega}^{AB}{}_t$, and $\Omega^{AB}{}_t$, and so they are first-class
constraints according to the above-mentioned expression for $H_1$. Further, the
spatial contractions in $\tilde{{\cal H}}_\perp, \tilde{{\cal H}}_d$, and $\tilde
{{\cal H}}_{AB}$ can be replaced by space-time contractions by including $\phi_{AB}
{}^t$ into the original expressions, leading to
\begin{equation}\label{Hp}
{\cal H}_\perp=\phi_A D_\perp\xi^A+\phi_{AB}{}^\mu{\cal F}^{AB}{}_{\perp\mu},
\end{equation}
\begin{equation}
{\cal H}_d=\phi_A D_d\xi^A+\phi_{AB}{}^\mu{\cal F}^{AB}{}_{d\mu},
\end{equation}
\begin{equation}\label{HAB}
{\cal H}_{AB}=-D_\mu\pi_{AB}{}^\mu-\pi_{[A}\,\xi_{B]},
\end{equation}
all of which are first-class constraints again. They are called the lapse, shift,
and dS constraints, respectively. As will be seen, these constraints represent the
normal/tangential diffeomorphism invariance, and dS invariance of the system.

Generally, the symmetry transformation of ${\cal P}$ is defined by
\begin{equation}\label{xg}
\xi^A\rightarrow g^A{}_B\,\phi_*\xi^B,
\end{equation}
\begin{equation}
\Omega^A{}_{B\mu}\rightarrow g^A{}_C\,\phi_*\Omega^C{}_{D\mu}
\,(g^{-1})^D{}_B+g^A{}_C\,\partial_\mu(g^{-1})^C{}_B,
\end{equation}
\begin{equation}
\pi_A\rightarrow \phi_*\pi_B\,(g^{-1})^B{}_A \det(\partial\phi_*
x^\nu/\partial x^\sigma),
\end{equation}
\begin{equation}\label{pABg}
\pi_{AB}{}^\mu\rightarrow \phi_*\pi_{CD}{}^\mu\,(g^{-1})^C{}_A
\,(g^{-1})^D{}_B \det(\partial\phi_*x^\nu/\partial x^\sigma),
\end{equation}
where $g^A{}_B$ is an $SO(1,4)$-valued function, and $\phi_*$ is the pushforward
by a diffeomorphism transformation $\phi$. Vary $g^A{}_B$ and $\phi$ to
give the one-parameter local groups $(g_\lambda)^A{}_B$ and $\phi_\lambda$
with the parameter $\lambda$. Differentiation of Eqs.
(\ref{xg})--(\ref{pABg}) with respect to $\lambda$ gives rise to the
infinitesimal transformation:
\begin{equation}\label{dx}
\delta\xi^A=A^A{}_B \xi^B-L_v\xi^A,
\end{equation}
\begin{equation}
\delta\Omega^A{}_{B\mu}=-D_\mu A^A{}_B-L_v\Omega^A{}_{B\mu},
\end{equation}
\begin{equation}
\delta\pi_A=-\pi_B A^B{}_A-L_v\pi_A-\pi_A\partial_\nu v^\nu,
\end{equation}
\[
\delta\pi_{AB}{}^\mu=-\pi_{CB}{}^\mu A^C{}_A-\pi_{AC}{}^\mu A^C{}_B
\]
\begin{equation}\label{dpAB}
-L_v\pi_{AB}{}^\mu-\pi_{AB}{}^\mu\partial_\nu v^\nu,
\end{equation}
where $A^A{}_B=\partial/\partial\lambda|_{\lambda=0}\,(g_\lambda)^A{}_B$, $v|_x=\partial
/\partial\lambda|_{\lambda=0}\,(\phi_\lambda x)$, and $L_v$ is the Lie derivative along $v$,
e.g., $L_v\pi_{AB}{}^\mu=v^\nu\partial_\nu\pi_{AB}{}^\mu-\pi_{AB}{}^\nu\partial_\nu v^\mu$.
In the above transformation, putting $v=0$ yields the dS transformation $\delta_A$, while
putting $A^A{}_B=-\Omega^A{}_{B\mu}v^\mu$ yields the dS-invariant diffeomorphism $\delta_v$
\cite{Lu16}.
These transformations can be generated by the first-class constraints (\ref{Hp})--(\ref{HAB})
in the following way. Firstly, a function $F$ of the constrained phase space is said to be
generating a symmetry, if its Hamiltonian vector field $\chi_F\equiv\{F,\cdot\}$ generates
a symmetry. Secondly, define the distributional quantities corresponding to the constraints
(\ref{Hp})--(\ref{HAB}): $H(N)=\int_\Sigma d^3x\,N{\cal H}_\perp$, $H(\vec{N})=\int_\Sigma
d^3x\,N^a{\cal H}_a$, and $H(\Omega)=\int_\Sigma d^3x\,\Omega^{AB}{\cal H}_{AB}$,
where $\vec{N}$ stands for $N^a$, and $\Omega=\Omega^A{}_B$ is an $so(1,4)$-valued function.
It follows that
\begin{equation}\label{HN}
\{H(N),\cdot\}\approx \delta_{Nn},\ \
\{H(\vec{N}),\cdot\}\approx \delta_{\vec{N}},\ \
\{H(\Omega),\cdot\}\approx \delta_{\Omega},
\end{equation}
where $Nn$ is short for $Nn^\mu$. They show that the distributional lapse/shift
and dS constraints generate the normal/tangential diffeomorphism and dS transformations
of the constraint surface, respectively. Also, notice that the inclusion of $\phi_{AB}
{}^t$ in Eqs. (\ref{Hp})--(\ref{HAB}) is necessary for the validity of Eq.
(\ref{HN}) acting on $\Omega^{AB}{}_t$.

\subsection{Second-class constraints and Dirac bracket}
According to Dirac's quantization procedure, the constraints are solved after they
are quantized, resulting in a physical Hilbert space. When acting on this Hilbert space,
the constraint operators as well as their commutators give zero, and hence the constraint
algebra should be closed under the Poisson/Lie bracket \cite{Dirac,Thiemann}.

The constraints of EC gravity (\ref{pcA})--(\ref{pcB}) and (\ref{sc}) can be recombined
into the first-class ${\cal H}_\perp,{\cal H}_d,{\cal H}_{AB}$, $\phi_{AB}{}^t$ and the
second-class $\phi_{AB}{}^a$. Because of the existence of second-class constraints, they
do not form a closed algebra. To get rid of the second-class constraints, first find out
the independent components of them, which would not become first class after some
combinations. Let us assume that $C\equiv C^{AB}{}_c\,\phi_{AB}{}^c$ is first class,
then $\{C,\phi_{AB}{}^a\}\approx 0$. In virtue of $\{\phi_{AB}{}^a(\vec{x}), \phi_{CD}{}
^b(\vec{z})\}=\{(2\,\partial^2{\cal L}/\partial{\cal F}^{CD}{}_{tb}\,\partial D_a\xi^{[A})
\,\xi_{B]}-(2\,\partial^2{\cal L}/\partial{\cal F}^{AB}{}_{ta}\,\partial D_b\xi^{[C})\,
\xi_{D]}\}\,\delta(\vec{x}-\vec{z})$,
one have $C^{AB}{}_c=(M^{ab}{}_c+M^{[a}\delta^{b]}{}_c)(D_a\xi^A)(D_b\xi^B)$, where
$M^{ab}{}_c$ is an arbitrary tensor antisymmetric in the $ab$ indices, and $M^a=M^{ca}
{}_c$. Then the independent second-class constraints can be taken by $\phi_I=\phi_{td}
{}^b,M^{[a}\delta^{b]}{}_c\,\phi_{ab}{}^c,\phi_{\mu4}{}^b$, where $\phi_{\mu\nu}{}^c
\equiv\phi_{AB}{}^c(D_\mu\xi^A)(D_\nu\xi^B)$ and $\phi_{\mu4}{}^b\equiv\phi_{AB}{}^b
(D_\mu\xi^A)\,\xi^B$. Equivalently, we may set $\phi_I=\phi_{td}{}^b,\ \phi_{ad}{}^a,
\ \phi_{\mu4}{}^b$, where the arbitrary $M^a$ is eliminated.

Secondly, modify the Poisson bracket into the Dirac bracket as below:
\begin{equation}
\{F,F'\}_D=\{F,F'\}+\int_\Sigma d^3x\,C^{IJ}(\vec{x})
\{\phi_I(\vec{x}),F\}\{\phi_J(\vec{x}),F'\},
\end{equation}
where $F,F'$ are functions of the phase space ${\cal P}$, and $C^{IJ}(\vec{x})$
is antisymmetric, subject to $C^{IJ}(\vec{x})\{\phi_J(\vec{x}), \phi_K
(\vec{z})\}\approx\delta^I{}_K\delta(\vec{x}-\vec{z})$. The definition is a
generalization of the original Dirac bracket \cite{Dirac} from finite degrees
to infinite degrees of freedom. For any first-class constraint $\phi$, $\{\phi,\cdot\}
_D\approx\{\phi,\cdot\}$, and thus it is still first class under the new bracket. On the
other hand, $\{\phi_I,\cdot\}_D\approx 0$, and thus the second-class constraints become
first class now. To conclude, as long as the $C^{IJ}(\vec{x})$ is solved, all the
second-class constraints disappear, then the constraint algebra becomes closed.
For the solution of $C^{IJ}(\vec{x})$, its existence is supported by the independence of
the components of $\phi_I$. Specifically, the solution is
\begin{equation}
C^{td}{}_b{}^{\mu4}{}_a=(l^{-2}/\sqrt{-g})\cdot(\delta^d{}_b\,\delta^\mu{}_a-2\,
\delta^d{}_a\,\delta^\mu{}_b),
\end{equation}
\begin{equation}
C^{ad}{}_a{}^{\mu4}{}_b=(l^{-2}/\sqrt{-g})\,\delta^d{}_b\,\delta^\mu{}_t,
\end{equation}
and other independent components being equal to zero. To rewrite the Dirac bracket
in a manifestly dS-invariant way, define $C^{AB}{}_a{}^{CD}{}_b\,\phi_{AB}{}
^a\,\phi_{CD}{}^b\equiv C^{td}{}_a{}^{\mu4}{}_b\,\phi_{td}{}^a\,\phi_{\mu4}{}^b
+C^{ad}{}_a{}^{\mu4}{}_b\cdot\phi_{ad}{}^a\,\phi_{\mu4}{}^b$, and $C^{AB}{}_a{}
^{CD}{}_b=-C^{BA}{}_a{}^{CD}{}_b=-C^{AB}{}_a{}^{DC}{}_b$. Then one can derive
\begin{eqnarray}
C^{AB}{}_a{}^{CD}{}_b=(l^{-2}/\sqrt{-g})\cdot\left(D_a\xi^{[A}\,D_b\xi^{B]}\,D_t\xi^{[C}
\cdot\xi^{D]}+\right.\nonumber\\
\left.D_t\xi^{[A}\,D_a\xi^{B]}\,D_b\xi^{[C}\cdot\xi^{D]}+\right.\nonumber\\
\left.2\,D_b\xi^{[A}\,D_t\xi^{B]}\,D_a\xi^{[C}\cdot\xi^{D]}\right),
\end{eqnarray}
\begin{equation}\label{DBdS}
\{F,F'\}_D=\{F,F'\}+\int_\Sigma d^3x\left(\,C^{AB}{}_a{}^{CD}{}_b(\vec{x})
\{\phi_{AB}{}^a(\vec{x}),F\}\{\phi_{CD}{}^b(\vec{x}),F'\}
-F\leftrightarrow F'\,\right),
\end{equation}
which are dS covariant as expected.

\section{Remarks}
The present work contributes to the dS-covariant generalization of the Hamiltonian
EC gravity. In the Lorentz gauges, our results coincide with those in the
Lorentz-covariant formalism \cite{Stefano,Nikolic}. The physical effect associated
with our formalism lies in the dS spin, which appears in the gravitational
field equation (\ref{ECeq}). For the geometrical part, the dS spin contains the
torsion and the cosmological constant. For the material part, it is a 5d generalization
of the Lorentz spin, and should be analyzed in the context of a quantum theory, as well
as its semiclassical limit.

The linearly realized formulation also helps us to distinguish translation and
diffeomorphism. In this formulation, they are different by definition, with
different features as follows. Firstly, the diffeomorphism symmetry is a
fundamental symmetry, which does not correspond to any conservation law directly.
In fact, the energy-momentum conservation results from both the translation and
diffeomorphism invariance, and likewise, the angular momentum conservation
results from both the Lorentz and diffeomorphism invariance \cite{Lu17}.
Secondly, the distributional dS constraint satisfies $\{H(\Omega),
H(\Omega')\}=-H([\Omega,\Omega'])$, indicating that the localization of the dS
group does not deform the dS algebra, including the translation algebra embedded
in it. On the other hand, the diffeomorphism algebra deforms the translation
algebra, see, e.g. Ref. \cite{Baekler}.

\section*{Acknowledgments}
I would like to thank Profs. S.-D. Liang and Z.-B. Li for their abiding help. Also,
special thanks are owe to Prof. F. W. Hehl who sent me the valuable book
\cite{Hehl13} in time, so that the work \cite{Nikolic} has reached me in time.

\end{document}